\documentclass[aps,twocolumn,amsmath,showpacs,amssymb,superscriptaddress,nofootinbib]{revtex4-1}

\usepackage{graphicx}% Include figure files
\usepackage{dcolumn}% Align table columns on decimal point
\usepackage{bm}% bold math

\usepackage{hyperref}

\usepackage{color}
\usepackage{amsfonts}
\usepackage{amssymb}
\usepackage{amsmath}

\newcommand{\gton}{\mathrel{\lower.9ex \hbox{$\stackrel{\displaystyle
>}{\sim}$}}}
\newcommand{\lton}{\mathrel{\lower.9ex \hbox{$\stackrel{\displaystyle
<}{\sim}$}}}

\begin{document}

\title{Resonance decay dynamics and their effects on $p_T$-spectra of pions in heavy-ion collisions}

\author{Pok Man Lo}
\affiliation{Institute of Theoretical Physics, University of Wroclaw,
PL-50204 Wroc\l aw, Poland}
%\affiliation{Extreme Matter Institute EMMI, GSI, D-64291 Darmstadt, Germany}

\begin{abstract}
	We examine the influence of resonance decay dynamics on momentum spectra of pions in heavy-ion collisions. 
	Taking the decay $\omega \rightarrow 3 \pi$ as an example, we demonstrate how details of the decay matrix element can modify the physical observables.
  Such a dynamical effect is commonly neglected in statistical models. To remedy the situation,
	we formulate a theoretical framework for incorporating hadron dynamics into the analysis, which can be straightforwardly
	extended to describe general N-body decays.
\end{abstract}

%\pacs{25.75.-q, 25.75.Ld, 12.38.Mh, 24.10.Nz}

\maketitle

\section{introduction}

The problem posed by heavy-ion collisions is to deduce physical properties of the created hadronic matter 
based on information of the observed particles. 
Since most of the particles detected are connected with the system after the freeze-out stage, 
precise modeling at different levels is required to reconstruct the cooling history of the originally produced hot and dense medium.

Pion production is a dominant feature in heavy-ion collisions.
The experimental data on momentum distributions of pions and other identified particles 
present a handle for discerning particles of different momenta.
In many cases the hadronic spectra are well reproduced by simple thermodynamical fits~\cite{Sollfrank:1990qz}. 
However the situation is more complicated for pions. 
In particular the data exhibit an unexpected enhancement of pions at low transverse momentum ($p_T$)~\cite{Abelev:2013vea}.
Presumably multiple mechanisms contribute to the observed spectrum. 
These include collective flow~\cite{Huovinen:2001cy}, resonance decay~\cite{Sollfrank:1990qz}, influence of the medium~\cite{Shuryak:2002kd,Rapp:2003ar}, and non-equilibrium effects~\cite{Petran:2013lja,Begun:2014rsa,Begun:2015ifa}. 

Therefore it is no longer sufficient to 
perform data fitting within the framework of a single mechanism. 
Instead a detailed examination of each effect is required,
with its consistency with known hadron physics and symmetries of QCD inspected.
Only then one can isolate the (possibly dominant) effect from the thermal medium and non-equilibrium dynamics, 
and eventually arrive at an internally consistent picture for heavy ion collisions.

In this work we focus on the $p_T$-spectra of pions from resonance decays. 
It is known that a good description of resonances is essential for understanding the soft part of the spectrum. 
An extensive analysis based on statistical models was presented in Ref.~\cite{Sollfrank:1990qz}. However, in this and other studies~\cite{Chojnacki:2011hb, Torrieri:2004zz}, resonance decay dynamics has been neglected. 

The purpose of this paper is to formulate a theoretical framework for incorporating
hadron dynamics into the analysis, applicable to a general N-body decay.
The detail of this framework is discussed in Sec. 2. In Sec. 3, we apply the formalism to study the three-body decay of $\omega \rightarrow 3 \pi$. 
In Sec. 4 we present the conclusion.

\section{momentum distributions of decay particles with decay dynamics}

\subsection{differential phase space}

The first question to address is to determine the distribution $ d n_1^{\rm dec}/d^3 p_1 $ of a particular decay particle $1$ (in this case a pion) from a given distribution $ d n_{\rm res}/d^3 p_{\rm res} $ of the resonance. 
A detailed account of this problem is given in textbooks~\cite{textbook1, Byckling:1971vca}. The application in heavy-ion collisions is discussed in Ref.~\cite{Sollfrank:1990qz, Gorenstein:1987zm}. Here we briefly review the key steps of the calculation to establish our notations.

The pion momentum spectrum from resonance decay is given by

\begin{align}
	\label{eq:3p_spec}
	E_\pi \frac{d n^{\rm dec}_{\pi}}{d^3 p_\pi} = {\rm br} \times \int d^3 p_{\rm res} \,  \frac{d n_{\rm res}}{d^3 p_{\rm res}} \times E_\pi^\star \times {\rm dPS}(\vec{p}_\pi^{\,\star}). 
\end{align}

\noindent Here ${\rm br}$ is the suitable branching ratio for the decay. The differential phase space function ${\rm dPS}(\vec{p}_\pi^{\,\star})$ is a key quantity of this study and will be addressed in detail. Throughout this work, variables in the resonance rest frame are denoted by $\star$, while those without are in the rest frame of the medium. 
The momentum variable $\vec{p}^{\,\star}_{\pi}$ should be understood as a function of $\vec{p}_{\rm res}$ and $ \vec{p}_{\pi}$. The explicit expression is easily obtained by invoking Lorentz invariance of $p_\pi \cdot p_{\rm res}$, which dictates

\begin{align}
	\begin{split}
	E_\pi^\star &= \frac{ E_\pi E_{\rm res} - \vec{p}_\pi \cdot \vec{p}_{\rm res} }{ m_{\rm res} } \\
	p_\pi^\star &= \sqrt{ {E_\pi^\star}^2 - m_\pi^2 }.
	\end{split}
\end{align}

In this study we shall only consider resonances from a static thermal source. Hence we put

\begin{align}
	\frac{d n_{\rm res}}{d^3 p_{\rm res}} \rightarrow \frac{g_{\rm res}}{(2 \pi)^3}\frac{1}{e^{\beta E_{\rm res}}-1}
\end{align}

\noindent for a mesonic resonance of degeneracy $g_{\rm res}$ at finite temperature ($T = 155 \,$ MeV is chosen) and vanishing chemical potentials.

The function ${\rm dPS}$ involves an integral over the phase space of other decay particles and the decay matrix element in the resonance rest frame. For an N-body decay the general definition reads~\cite{Sollfrank:1990qz, textbook1, Byckling:1971vca, Kumar:1970cr, Becattini:2009sc}

\begin{align}
	\label{eq:dps}
	\begin{split}
		{\rm dPS}(\vec{p}_1^{\, \star}) &= \frac{1}{{\gamma_{\rm res}}} \frac{d \gamma_{\rm res} }{d^3 p_1^\star}   \\
	    &=  \frac{1}{{\gamma_{\rm res}}} \frac{1}{2 m_{\rm res}}  \frac{1}{(2 \pi)^3} \frac{1}{2 E_1^\star}  \times  \\
	&\int \frac{d^3 p_2^\star}{(2 \pi)^3} \frac{1}{2 E_2^\star}\frac{d^3 p_3^\star}{(2 \pi)^3} \frac{1}{2 E_3^\star} \cdots \frac{d^3 p_N^\star}{(2 \pi)^3}\frac{1}{2 E_N^\star} \times  \\
	& (2 \pi)^4 \delta^4(P-\sum_i p_i) \, \vert \Gamma_{{\rm res} \rightarrow 1 + 2 + \cdots + N } \vert^2 
	\end{split}
\end{align}

\noindent  The normalization of this function is chosen such that

\begin{align}
	\int d^3 p_1^\star \, ({\rm dPS}) = 1.
\end{align}

A common approximation made by thermodynamical models when calculating this quantity is the assumption of isotropic (structureless) decay \cite{Sollfrank:1990qz}. This amounts to replacing the decay matrix element $\Gamma$ with the identity $I$ and hence ${\rm dPS} \rightarrow {\rm dPS}^{(0)}$:

\begin{align}
	\begin{split}
		{\rm dPS}^{(0)} &= \frac{1}{\phi_N} \, \frac{d\phi_N }{d^3 p_1^\star}  \\
	&=  \frac{1}{\phi_N} \frac{1}{(2 \pi)^3} \frac{1}{2 E_1^\star} \times   \\
	&\int \frac{d^3 p_2^\star}{(2 \pi)^3} \frac{1}{2 E_2^\star}\frac{d^3 p_3^\star}{(2 \pi)^3} \frac{1}{2 E_3^\star} \cdots \frac{d^3 p_N^\star}{(2 \pi)^3}\frac{1}{2 E_N^\star} \times  \\
	& (2 \pi)^4 \delta^4(P-\sum_i p_i). 
	\end{split}
\end{align}

\noindent Here we have introduced the N-body Lorentz invariant phase space $\phi_N$:

\begin{align}
	\begin{split}
	\phi_N  &= \int d \phi_N \\
		&= \int \frac{d^3 p_1^\star}{(2 \pi)^3} \frac{1}{2 E_1^\star}\frac{d^3 p_2^\star}{(2 \pi)^3} \frac{1}{2 E_2^\star}\cdots \frac{d^3 p_N^\star}{(2 \pi)^3}\frac{1}{2 E_N^\star} \times  \\
	& (2 \pi)^4 \delta^4(P-\sum_i p_i). 
	\end{split}
\end{align}

        To clarify the physical meaning of the differential phase space function ${\rm dPS}$, we explicitly work out the cases for isotropic two- and three-body decay. Starting with the two-body case:

\begin{align}
	\begin{split}
		{\rm dPS}^{(0)} &= \frac{1}{\phi_2} \frac{1}{(2 \pi)^3} \frac{1}{2 E_1^\star} \times \int \frac{d^3 p_2^\star}{(2 \pi)^3} \frac{1}{2 E_2^\star} \times  \\
	& (2 \pi)^4 \delta(m_{\rm res}-E_1^\star-E_2^\star) \delta^3(\vec{p}_1^{\, \star} + \vec{p}_2^{\, \star}).
	\end{split}
\end{align}

\noindent The integrals can be explicitly worked out: 

\begin{align}
	\begin{split}
	I_2 &= \frac{1}{(2 \pi)^3} \frac{1}{2 E_1^\star} \times \int \frac{d^3 p_2^\star}{(2 \pi)^3} \frac{1}{2 E_2^\star} \times  \\
        &(2 \pi)^4 \delta(m_{\rm res}-E_1^\star-E_2^\star) \delta^3(\vec{p}_1^{\, \star} + \vec{p}_2^{\, \star})  \\
	&= \frac{1}{4 m_{\rm res} q} \frac{1}{(2 \pi)^2} \delta(p_1^\star - q) ,
	\end{split}
\end{align}

\begin{align}
	\begin{split}
	\phi_2(m_{\rm res}^2, m_1^2, m_2^2) &= \frac{1}{8 \pi m_{\rm res}^2} \sqrt{ \lambda(m_{\rm res}^2, m_1^2, m_2^2) }  \\
	&= \frac{q}{4 \pi m_{\rm res}},
	\end{split}
\end{align}

\noindent where $\lambda(x,y,z)$ is the K{\"a}ll{\'e}n triangle function~\cite{Byckling:1971vca}

\begin{align}
\lambda(x,y,z) = x^2 + y^2 + z^2 - 2 x y - 2 x z - 2 y z.
\end{align}

\noindent and $q$ is the three-momentum of the decay particle in the resonance rest frame

\begin{align}
	\begin{split}
	q &= \frac{1}{2} \sqrt{ m_{\rm res}^2 } \times  \\
	  &\sqrt{ 1 - \frac{(m_1 + m_2)^2}{m_{\rm res}^2} } \sqrt{ 1 - \frac{(m_1 - m_2)^2}{m_{\rm res}^2} }. 
	\end{split}
\end{align}

\noindent Finally we arrive at the well-known result

\begin{align}
	{\rm dPS}^{(0)} = \frac{1}{4 \pi q^2} \delta(p_\pi^\star -q),
\end{align}

\noindent which may be alternatively obtained by inspection of the quantity $dN^{\rm dec}_1/d^3 p_1^\star$ for a spherically symmetric two-body decay~\cite{Sollfrank:1990qz,Gorenstein:1987zm}.

The generalization to the case of three-body decay is straightforward:

\begin{align}
	\begin{split}
		{\rm dPS}^{(0)} &= \frac{1}{\phi_3} \frac{1}{(2 \pi)^3} \frac{1}{2 E_1^\star} \times \\
	&\int \frac{d^3 p_2^\star}{(2 \pi)^3} \frac{1}{2 E_2^\star}  \frac{d^3 p_3^\star}{(2 \pi)^3} \frac{1}{2 E_3^\star} \times   \\
	& (2 \pi) \delta(m_{\rm res}-E_1^\star-E_2^\star-E_3^\star) \\
	&(2 \pi)^3 \delta^3(\vec{p}_1^{\, \star} + \vec{p}_2^{\, \star} + \vec{p}_3^{\, \star}).
	\end{split}
\end{align}

\noindent Where the integrals can again be explicitly calculated: 

\begin{align}
	\begin{split}
	I_3 &= \frac{1}{(2 \pi)^3} \frac{1}{2 E_1^\star} \times \\
	&\int \frac{d^3 p_2^\star}{(2 \pi)^3} \frac{1}{2 E_2^\star} \frac{d^3 p_3^\star}{(2 \pi)^3} \frac{1}{2 E_3^\star}  \times   \\
	&(2 \pi) \delta(m_{\rm res}-E_1^\star-E_2^\star-E_3^\star) \\
	&(2 \pi)^3 \delta^3(\vec{p}_1^{\, \star} + \vec{p}_2^{\, \star} + \vec{p}_3^{\, \star}) \\
	&= \frac{1}{(2 \pi)^3} \frac{1}{2 E_1^\star} \times \\
	&\frac{1}{8 \pi (P-p_1)^2} \sqrt{ \lambda((P-p_1)^2, m_2^2, m_3^2) }
	\end{split}
\end{align}

\begin{align}
	\begin{split}
	\phi_3(s) &= \frac{1}{16 \pi^2} \frac{1}{s} \times  \\
	&\int^{(\sqrt{s}-m_1)^2}_{(m_2+m_3)^2} d s^\prime \sqrt{\lambda(s,s^\prime, m_1^2)} \times \\
	&\phi_2(s^\prime, m_2^2, m_3^2). 
	\end{split}
\end{align}

\noindent Finally the differential phase space for isotropic three-body decay reads

\begin{align}
	{\rm dPS}^{(0)} = \frac{I_3}{\phi_3(s = m_{\rm res}^2)},
\end{align}

\noindent matching the result in Ref.~\cite{Sollfrank:1990qz}.

	The assumption of isotropic decay can be justified in some cases when the matrix element is a scalar (e.g. $\sigma \rightarrow \pi\pi$ decay via $\mathcal{L}_{\rm int} = -g \, \sigma \pi \pi$) or depends only on $s = P^2$. However, as we shall demonstrate in the example of the 3-body decay $\omega \rightarrow 3 \pi$, this approximation is problematic especially for soft pions.

\section{case study: $\omega \rightarrow 3 \pi$}

\subsection{${\rm dPS}$ and $p_T$-spectra}

Multiple hadron models \cite{Klingl:1996by, Leupold:2008bp, Niecknig:2012sj, Danilkin:2014cra} are available to describe the decay of $\omega$-meson to three pions. 
The mechanisms involved are the Gell-Mann, Sharp and Wagner (GSW) process~\cite{GellMann:1962jt, Wess:1971yu} ($\omega \rightarrow \rho \pi \rightarrow \pi \pi \pi$) and possibly a direct process~\cite{Klingl:1996by}.
Here we employ the model of Ref.~\cite{Danilkin:2014cra} which is a dispersive study based on the isobar decomposition and subenergy unitarity.
Accordingly, the decay matrix element is given by

\begin{align}
	\label{eq:matrix_element}
  \vert \Gamma_{\omega \rightarrow 3\pi} \vert^2 = \mathcal{P} \, \vert C_{V \rightarrow 123} \vert^2
\end{align}

\noindent where

\begin{align}
	\begin{split}
    \mathcal{P} &= -\frac{1}{3} \epsilon_{\mu \nu \alpha \beta} \epsilon_{a b c d} P^\mu \, p_1^\nu \, p_2^\alpha \,  P^a \, p_1^b \, p_2^c \, g^{\beta d} \\
		&= \frac{1}{12} \times \left( s_{12} s_{23} s_{13} - m_\pi^2 (m_{\rm res}^2 - m_\pi^2 )^2 \right),
	\end{split}
\end{align}

\noindent and

\begin{align}
	s_{ij} = (p_i + p_j)^2,
\end{align}

\noindent all subjected to the kinematic constraint

\begin{align}
	P^2 = m_{\rm res}^2 = s_{12} + s_{23} + s_{13} - m_1^2 - m_2^2 - m_3^2.
\end{align}

The factor $\mathcal{P}$ due to anomalous coupling, which is common to all models of the decay, dominates the properties of the matrix element. 
On the other hand, differences among models are limited to the different recipe for the amplitude function $C_{V \rightarrow 123}$.
We have numerically confirmed that different choices of the latter only lead to minimal changes to our subsequent results\footnote{Deviation among models becomes appreciable when we dial up the meson width.}.
The detail of the amplitude function employed in this work is given in Ref.~\cite{Danilkin:2014cra} and the expression is reproduced here for convenience:

\begin{align}
	\begin{split}
    \vert C_{V \rightarrow 123} \vert^2 &= \vert  \mathcal{N} \vert^2 \, ( 1 + 2 \alpha z + 2 \beta z^{3/2} \sin (3 \theta) \\ + 2 \gamma z^2 +
    &2 \delta z^{5/2} \sin (3 \theta)),
	\end{split}
\end{align}

\noindent where

\begin{align}
	\begin{split}
    \sqrt{z} \, \cos (\theta) &= \frac{\sqrt{3} \, (s_{23}-s_{13})}{2 \,  m_{\rm res}(m_{\rm res}- 3 \,  m_\pi)} \\
    \sqrt{z} \, \sin (\theta) &= \frac{\sqrt{3} \, (s_c-s_{12})}{2 \, m_{\rm res}(m_{\rm res}- 3 \, m_\pi}) \\
    s_c &= \frac{1}{3} \, (m_{\rm res}^2 + 3 m_\pi^2),
	\end{split}
\end{align}

\noindent with the model parameters

\begin{align}
	\begin{split}
    \alpha &= 0.083 \\
    \beta &= 0.022 \\
    \gamma &= 0.001 \\
    \delta &= 0.014. 
	\end{split}
\end{align}

\noindent The normalization $\mathcal{N}$ is chosen such that the integrated width matches the experimental value. 

The decay matrix element~\eqref{eq:matrix_element} is a function of phase space variables and thus cannot be pulled out of the integral in Eq.~\eqref{eq:dps}. In fact, the integration over the phase space of other decay particles is equivalent to the integration over the region of Dalitz decay:

\begin{align}
	\int d \phi_3 \, \cdots =  \frac{1}{128 \pi^3 M^2} \int_{\rm Dalitz} d s_{12} \, d s_{23} \, \cdots.
\end{align}

\begin{figure}[ht!]
	\centering
 \includegraphics[width=0.497\textwidth]{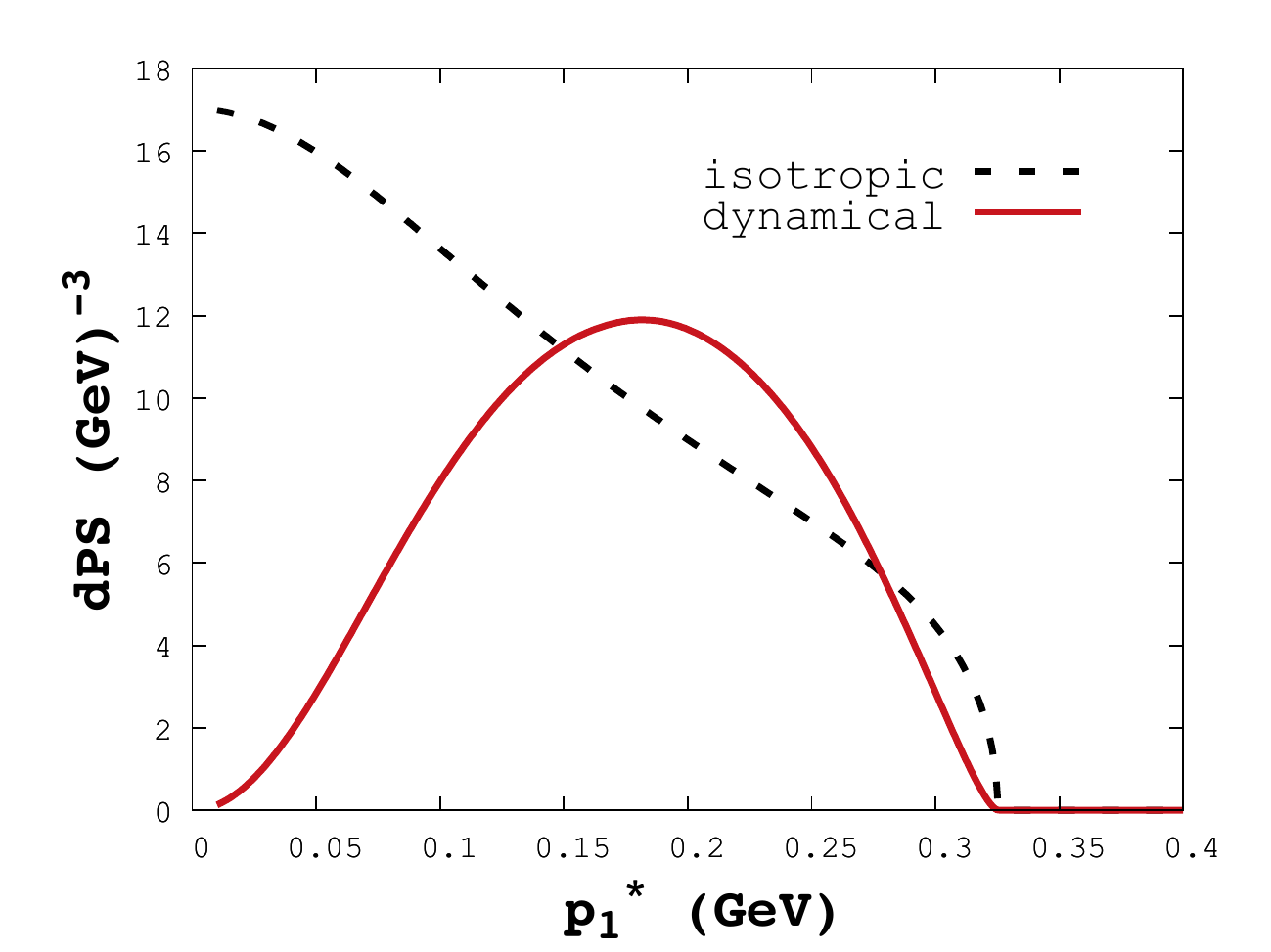}
	\caption{Differential phase space function ${\rm {\rm dPS}}$ calculated by Eq.~\eqref{eq:dps}
	for the decay $\omega \rightarrow 3 \pi$. Solid line corresponds to calculation with the full matrix element~\eqref{eq:matrix_element} while dashed line corresponds to result from the isotropic approximation.}
	\label{fig:one}
\end{figure}

\begin{figure*}[ht!]
	\centering
 \includegraphics[width=0.497\textwidth]{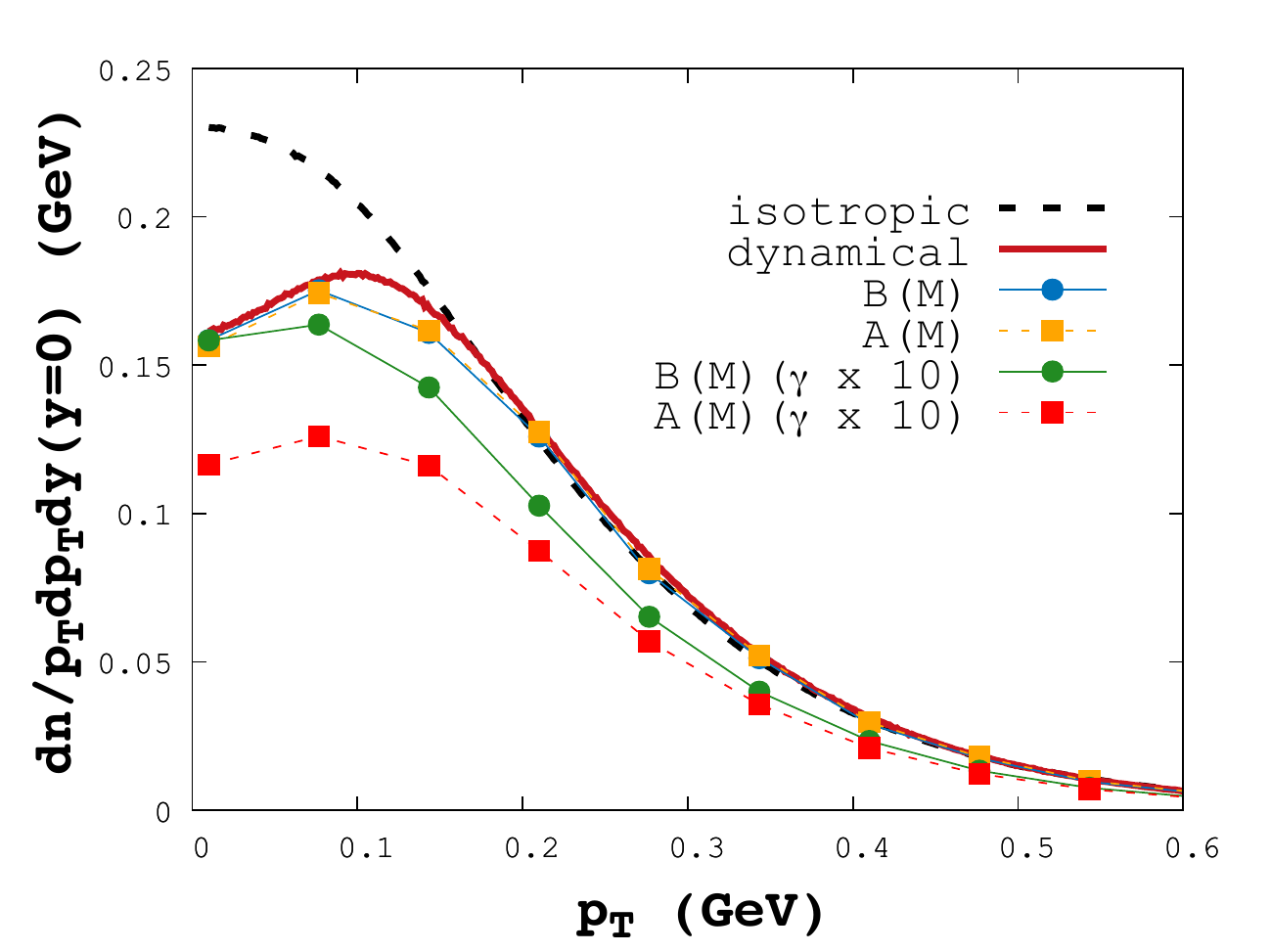}
 \includegraphics[width=0.497\textwidth]{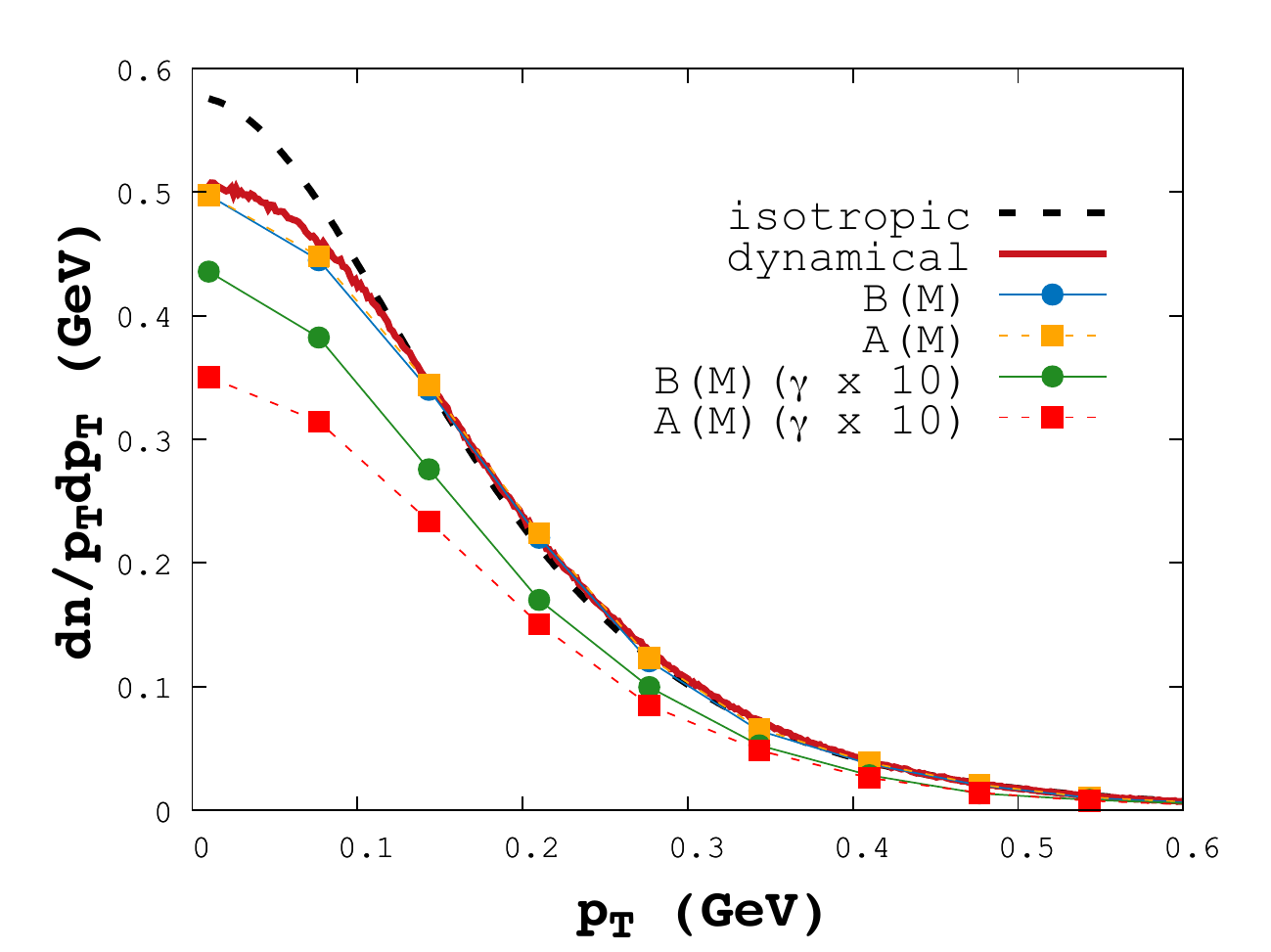}
	\caption{The mid-rapidity (left) and rapidity-integrated (right) $p_T$-spectra of decay pions from $\omega$-meson ( static source, $T= 155\,$ MeV ) decay, calculated for full decay matrix element in Eq.~\eqref{eq:matrix_element} and for the isotropic case. Also shown (as points) are the corresponding results including the width of the $\omega$-meson via the S-matrix approach calculated from Eq.~\eqref{eq:buspec}, for the case of physical resonance width and the case in which the coupling constant is scaled by a factor of 10.}
 \label{fig:two}
\end{figure*}

\noindent To examine the effects of decay dynamics on ${\rm dPS}$, we numerically compute the integral in Eq.~\eqref{eq:dps} together with the matrix element in Eq.~\eqref{eq:matrix_element}. Explicitly, the integral reads

\begin{align}
	\begin{split}
		{\rm dPS} &= \frac{1}{2 m_{\rm res} {\gamma_{\omega \rightarrow 3\pi}}} \frac{1}{(2 \pi)^3} \frac{1}{2 E_1^\star} \times \\
	&\int \frac{d^3 p_2^\star}{(2 \pi)^3} \frac{1}{2 E_2^\star}  \frac{d^3 p_3^\star}{(2 \pi)^3} \frac{1}{2 E_3^\star}  \times  \\
	& (2 \pi) \delta(m_{\rm res}-E_1^\star-E_2^\star-E_3^\star) \\
	&(2 \pi)^3 \delta^3(\vec{p}_1^{\, \star} + \vec{p}_2^{\, \star} + \vec{p}_3^{\, \star}) \times {\vert} \, \Gamma_{\omega \rightarrow 3\pi} (s_{ij})\, {\vert}^2 
	\end{split}
\end{align}

\noindent with

\begin{align}
	\begin{split}
    \label{eq:width}
		\gamma_{\omega \rightarrow 3 \pi} &= \frac{1}{2 m_{\rm res}} \int d \phi_3 \, \vert \Gamma_{\omega \rightarrow 3\pi} \vert^2. 
	\end{split}
\end{align}

\noindent In this implementation the integral in Eq.~\eqref{eq:width} for $\gamma_{\omega \rightarrow 3 \pi}$ is by construction given by the experimental value $\gamma_{\rm exp} = (7.57 \pm 0.13)$ MeV~\cite{Olive:2016xmw}. The result for the ${\rm dPS}$ function is shown in Fig.~\ref{fig:one}. The key observation is that the full ${\rm dPS}$ function is substantially suppressed at low momenta. This is a direct consequence of the factor $\mathcal{P}$ in the decay matrix element. Note that both functions are normalized to unity when an integration over $\int d^3 p_1^\star $ is performed. Furthermore, the cut at $p_1^\star \approx 0.33 $ GeV simply reflects the kinematical situation where all three decay particles are collinear, with particles 1 going one way and the others going the opposite.

Next we study the influence of dynamics on momentum distributions.
To construct the conventional $p_T$-spectra studied in experiments, 
we perform an additional integration over the rapidity range on Eq.~\eqref{eq:3p_spec}.
At this stage, even the kinematic cuts from a specific experimental analysis can be easily implemented. This may be essential for a realistic comparison with data~\cite{Broniowski:2003ax}. For simplicity we shall skip it here and
consider only the mid-rapidity ($y = 0$) and rapidity-integrated $p_T$-spectra of pions from the decay of $\omega$-meson. These are
shown in Fig.~\ref{fig:two}. 

It is somewhat surprising that despite the essential differences in the ${\rm dPS}$ functions (Fig.~\ref{fig:one}), the deviations in $p_T$-spectra yielded by different treatments of decay dynamics are rather mild (Fig.~\ref{fig:two}). In both spectra we see that the correction from dynamics is limited to the low-$p_T$ region. Moreover deviation from the isotropic case is more visible in the mid-rapidity spectrum than in the integrated one. This is expected as features of ${\rm dPS}$, and hence the influence of dynamics, will be washed out when all the momentum variables are integrated over.

Previous study~\cite{Sollfrank:1990qz} suggests that the $\omega$-meson is one of the major sources of low-$p_T$ pions. In this work, we find that imposing the correct decay dynamics leads to a reduced contribution. More unexpected effects can come from other resonances and other types of decay. Further work is required to revise the resonance decay contribution to various physical observables based on the input of robust hadron physics.

\subsection{effects of finite width}

Another consequence of hadron dynamics is the existence of resonance widths. These can be systematically included using the S-matrix formalism of Dashen, Ma and Bernstein~\cite{Dashen:1969ep}. 
In this model, the Feynman amplitude for $3 \pi \rightarrow 3 \pi$ scattering can be constructed from the decay matrix element
in Eq.~\eqref{eq:matrix_element} via an S-channel resonance exchange process

\begin{align}
	\label{eq:model}
	\begin{split}
		i \mathcal{M} &= \frac{-i \, \left\vert \Gamma \right\vert^2 / {\rm br}}{M^2 - \bar{m}_{\rm res}^2 + i M \gamma_{\rm tot}} \\
		\gamma_{\omega \rightarrow 3 \pi} &= \frac{1}{2 M} \int d \phi_3 \, \vert \Gamma \vert^2 \\	
		\gamma_{\rm tot} &\approx \gamma_{\omega \rightarrow 3 \pi}/{\rm br},
	\end{split}
\end{align}

\noindent where $M$ is the invariant mass, $\bar{m}_{\rm res} = 0.783$ GeV is the pole mass of $\omega$-meson and ${\rm br} = 0.892$ is the branching ratio. In this model we do not explicitly calculate the other partial widths of $\omega$-meson, but simply prescribe a factor of $1/{\rm br}$ on $\gamma_{\omega \rightarrow 3 \pi}$ to obtain the total width $\gamma_{\rm tot}$.

With the model scattering amplitude, the generalized phase shift function $\mathcal{Q}(M)$ and the effective spectral function $B(M)$ can be computed as follows~\cite{How,Fornal:2009xc,Tayl,kappa}:

\begin{align}
	\label{eq:ps}
	\begin{split}
		\mathcal{Q}(M) &= \frac{1}{2} \,  {\rm Im} \, \left[ \ln{( 1 + \int d \phi_3 \, i \mathcal{M} )} \right] \\
		B(M) &= 2 \, \frac{d}{d M} \mathcal{Q}(M).
	\end{split}
\end{align}

\begin{figure*}[ht!]
	\centering
 \includegraphics[width=0.497\textwidth]{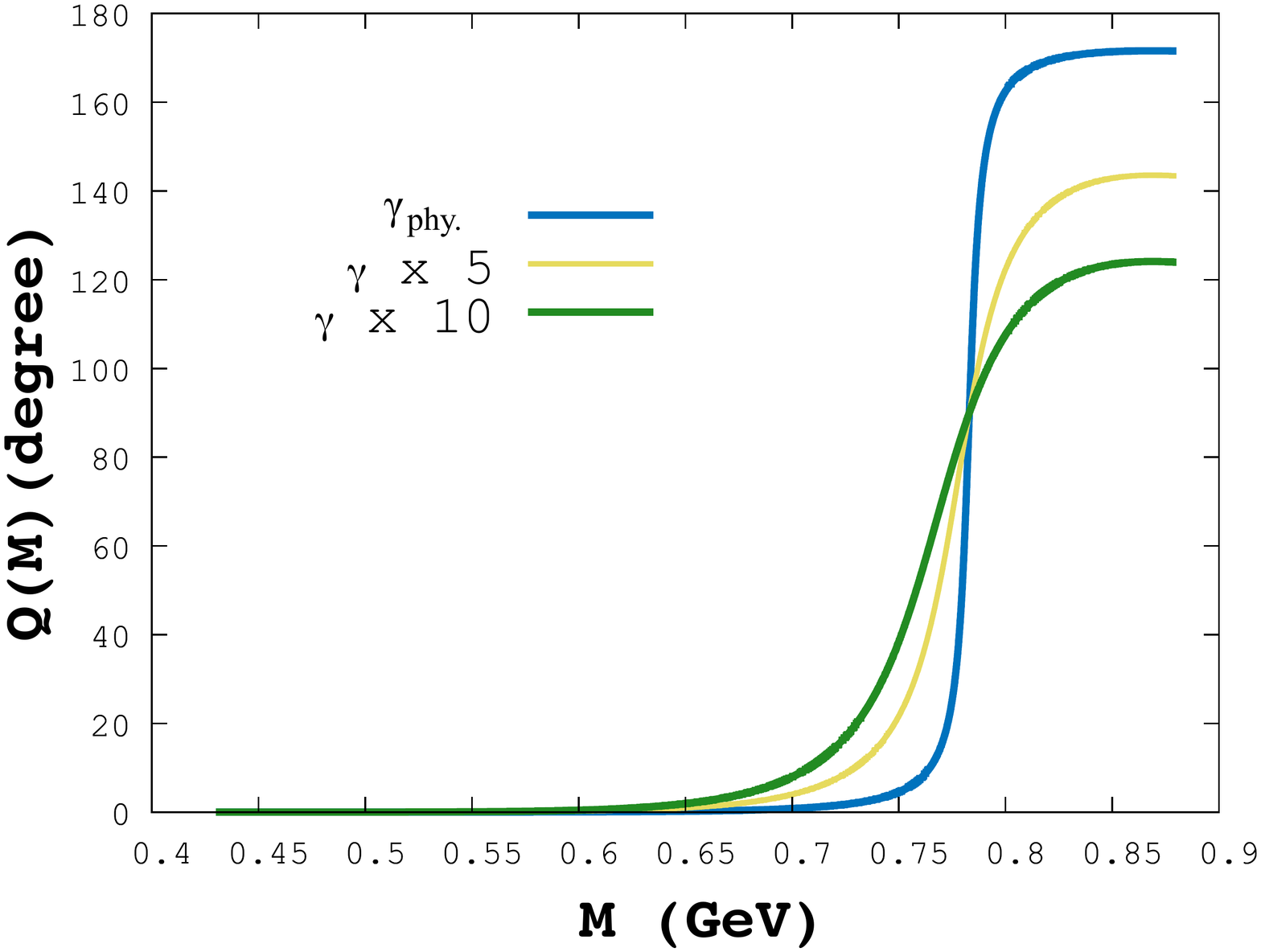}
 \includegraphics[width=0.497\textwidth]{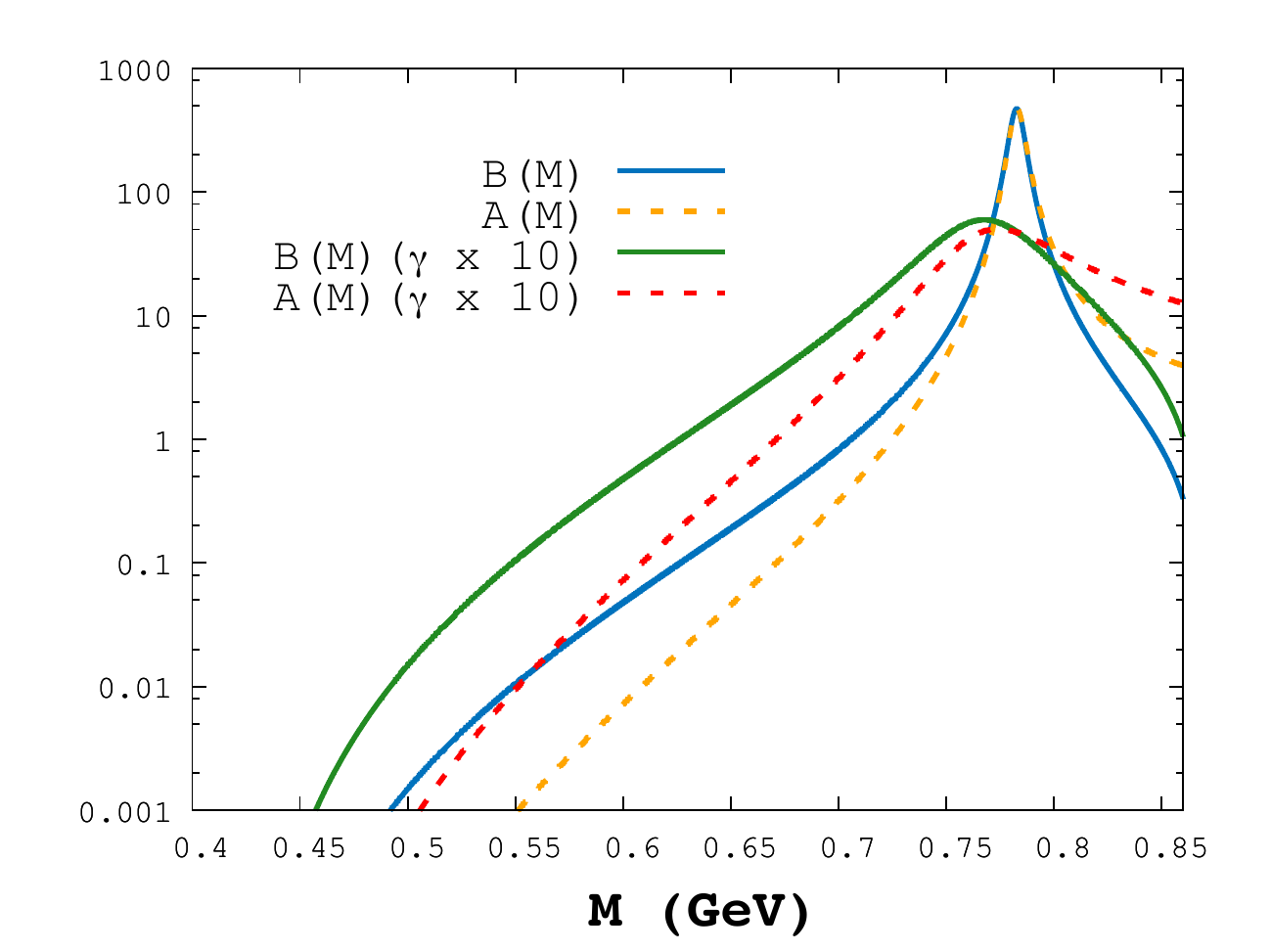}
	\caption{Generalized phase shift function $\mathcal{Q}$ and the effective spectral function $B$ computed from Eq.~\eqref{eq:ps} for the model amplitude~\eqref{eq:model}. The results are shown for the case of physical resonance width and the case in which the coupling constant is scaled by the designated factor.}
 \label{fig:three}
\end{figure*}

\noindent These functions are displayed in Fig.~\ref{fig:three}. Since the width of $\omega$-meson is small, the phase shift function indeed behaves like a theta-function $\pi \times \theta(M-\bar{m}_\omega)$, and the corresponding effective spectral function $B(M)$ is in practice well approximated by an energy-dependent Breit-Wigner function $A(M)$

\begin{align}
		A(M) &= -2 M \, \frac{\sin{2 \mathcal{Q}(M)}}{M^2-\bar{m}_\omega^2}.
\end{align}

For the momentum spectrum, the influence of resonance width enters via~\cite{rho-pt}

\begin{align}
	\label{eq:buspec}
	\begin{split}
		E_\pi \frac{d n^{\rm dec}_{\pi}}{d^3 p_\pi} &= {\rm br} \times \int^\Lambda \frac{d M}{2 \pi} \, B(M)  \times \\
		&\int_{m_{\rm res} \rightarrow M} d^3 p_{\rm res} \,  \frac{d n_{\rm res}}{d^3 p_{\rm res}} \times E_\pi^\star 
		\times {\rm dPS}(\vec{p}_\pi^{\,\star}).
	\end{split}
\end{align}

\noindent We perform analogous numerical integration on Eq.~\eqref{eq:buspec} as in the zero-width case. Here we pick $\Lambda = 0.88 \, {\rm GeV}$, which is how far we estimate the model given in Eq.~\eqref{eq:model} to hold. The results are shown in Fig.~\ref{fig:two}. Only moderate differences are found between this and the zero-width case. 
The vacuum width of $\omega$-meson is so narrow that the zero-width approximation is justified.

On the other hand, substantial broadening of the $\omega$-width in the medium is suggested by model studies~\cite{ Rapp:2000pe,Schneider:2001zt, Eletsky:2001bb, Lutz:2001mi, Riek:2004kx,Martell:2004gt}. To investigate the dependence on resonance width of our previous results, we simply scale the coupling constant in the matrix element in Eq.~\eqref{eq:matrix_element} by a factor of $10$. The resulting phase shift and the effective spectral functions are shown in Fig.~\ref{fig:three}.

In this particular model an increase of width leads not only to the broadening of the effective spectral functions, but also a reduction of their normalization although they are both normalized to unity in the limit of zero width. 
It also tends to reduce $\mathcal{Q}(M)$, and as $\gamma \rightarrow \infty$, the function will eventually approach the limit of $\pi/2$ at large invariant masses. As expected, this also leads to an overall drop in the magnitudes of the calculated $p_T$-spectra (Fig.~\ref{fig:two}).

Another important consequence of the broadening of resonance is the enhancement of low-$p_T$ pions from the use of the effective spectral function $B(M)$ over that from the use of $A(M)$. This can be traced to the greater value of $B(M)$ at low invariant masses, which translates into a larger contribution to the soft part of $p_T$-spectra due to the lesser Boltzmann suppression. A recent discussion of this effect is presented in Ref.~\cite{rho-pt} for the case of $\rho$-meson. In the case of $\omega$-meson, this effect becomes appreciable only if we increase the physical width by a factor of $10$.

\section{conclusion}

This study set out to investigate how details of hadron physics can modify heavy-ion collision observables.
To this end, we formulate a theoretical framework for incorporating resonance decay dynamics into the analysis.

As an application we consider the decay $\omega \rightarrow 3 \pi$, and 
find that imposing the anomalous coupling feature of the decay matrix element leads to a reduction of low-$p_T$ pions
compared to the structureless decay treatment.

In many statistical models, the isotropic decay approximation is adopted instead of the full dynamics.
The validity of this approximation has to be inspected case by case.
Since multiple mechanisms are at work to produce the observed $p_T$-spectra, it is necessary to perform a detailed examination 
of each effect based on existing knowledge of hadron physics.
In the current study, the finding of reduced low-$p_T$ pions from $\omega$-meson makes room for
other important effects such as higher order N-body decay, influence of thermal medium and non-equilibrium effects to
explain the unexpected enhancement of soft pions observed in the experiment.

It would be interesting to assess the effects of
other important features of the strong interaction 
on these observables. 
In particular, coupled-channel dynamics~\cite{Kaminski:2006yv,Guo:2010gx,Tornqvist:1995kr} and 
the existence of complex objects like hadronic molecules and other exotics~\cite{eric:exotica,Amsler:2004ps}.
Such research is currently underway.

\acknowledgments

I thank Francesco Becattini and Eric Swanson for carefully reading the manuscript and for giving constructive comments.
I am also grateful to Thomas Kl{\"a}hn for stimulating discussions and to Bengt Friman, Pasi Huovinen, Krzysztof Redlich and Chihiro Sasaki for the supportive collaboration.
This work was partly supported by the Polish National Science Center (NCN), under Maestro grant DEC-2013/10/A/ST2/00106 and by the Extreme Matter Institute EMMI, GSI.

\end{document}